\documentclass[msom,nonblindrev]{informs3_hide}
\DoubleSpacedXI % default for OR, MSOM
\usepackage{bbm,multirow,multicol,algorithm,algpseudocode}
\usepackage{cleveref}
\crefname{subsection}{subsection}{subsections}

\newcommand{\bI}{\mathbbm{1}}
\newcommand{\bE}{\mathbb{E}}

\newcommand{\ALG}{\mathsf{ALG}}

\newcommand{\DLP}{\mathsf{DLP}}
\newcommand{\nmax}{n_\mathrm{max}}
\newcommand{\nper}{n_\mathrm{per}}
\newcommand{\pcarry}{p_\mathrm{carry}}
\newcommand{\zsafety}{z_\mathrm{safety}}
\newcommand{\dist}{\mathsf{dist}}
\newcommand{\closest}{\mathsf{closest}}
\newcommand{\demand}{\mathsf{dem}}

\newcommand{\cunit}{c^{\mathrm{unit}}}
\newcommand{\cfix}{c^{\mathrm{fixed}}}

\usepackage[dvipsnames]{xcolor}

\usepackage{natbib}
 \bibpunct[, ]{(}{)}{,}{a}{}{,}%
 \def\bibfont{\small}%
\TheoremsNumberedThrough     % Preferred (Theorem 1, Lemma 1, Theorem 2)
\ECRepeatTheorems

\EquationsNumberedThrough    % Default: (1), (2), ...
\MANUSCRIPTNO{MSOM-22-390.R1} % When the article is logged in and DOI assigned to it,
\begin{document}
%%%%%%%%%%%%%%%%

%\setcounter{page}{1}
%\DOI{10.1287/xxxx.0000.0000}%


%\RUNTITLE{}

\TITLE{
Order-optimal Correlated Rounding for Fulfilling Multi-item E-commerce Orders
}

\ARTICLEAUTHORS{
\AUTHOR{Will Ma}
\AFF{Graduate School of Business, Columbia University, New York, NY 10027, \EMAIL{wm2428@gsb.columbia.edu}}
}

\ABSTRACT{

\SingleSpacedXI
We study a parsimonious correlated rounding problem motivated by e-commerce fulfillment.
In this problem, we are given a multi-item order of size $q$;
for each item $i=1,\ldots,q$, we are given the probability $u_{ki}$ with which it must be shipped from each Fulfillment Center (FC) $k=1,\ldots,K$, with $\sum_{k=1}^K u_{ki}=1$.
The goal is to randomly select a FC to ship each item following these marginal probabilities $u_{ki}$ (motivated by trying to satisfy long-run inventory flow), in a way that not many distinct FC's end up being used (motivated by reducing the number of boxes that need to be shipped).
In particular, the objective is to use each FC with probability at most $\alpha\cdot y_k$, where $y_k:=\max_{i=1,\ldots,q} u_{ki}$ is a lower bound on the probability with which FC $k$ must be used, and $\alpha\ge 1$ is the \textit{guarantee} to be made as small as possible.

This problem was originally introduced in Jasin and Sinha (2015), who present a rounding scheme that guarantees $\alpha\approx q/4$ given any $q$ marginal distributions.
Our main result is a rounding scheme that guarantees $\alpha=\ln(q)+1$, significantly improving the order-dependence on $q$ from linear to logarithmic.
We present another rounding scheme for sparse networks, that guarantees $\alpha=d$ if each item is stored in at most $d$ FC's.
We show these guarantees to be tight in terms of the dependence on $q$ or $d$.
Our schemes are simple and fast, based on two intuitive ideas---items wait for FC's to ``open'' following Poisson clocks, and observe them on ``dilated'' time scales.  The first idea positively correlates the FC's selected by items so that not many FC's are used; the second idea ensures that the marginal probability constraints are satisfied.

Interestingly, our main result implies a simple new correlated rounding scheme for the Set Cover randomized rounding problem.
Given a fractional set cover, it outputs an integral cover in which each set is selected with probability at most $\ln(q)+1$ times its fractional weight.
This improves the constant term from existing rounding schemes for Set Cover, which select each set with probability at least $\ln(q)+\omega(1)$ times its fractional weight.

Returning to the e-commerce application, it has been shown in Jasin and Sinha (2015) that results for this correlated rounding problem have direct implications for a dynamic fulfillment problem in which different multi-item orders arrive over time and there are inventory constraints.  We numerically test our rounding schemes under the dynamic fulfillment setups from Jasin and Sinha (2015) and find that they improve runtimes, shorten code, and robustly improve performance.  We make our code publicly available.

}

%\KEYWORDS{e-commerce fulfillment, randomized rounding, online algorithms}

\HISTORY{Full version of EC2023 paper}
%An initial version of this paper appeared in the Manufacturing \& Service Operations Management (MSOM) conference Special Interest Group (SIG) for Supply Chain Management, 2022.

\maketitle
%%%%%%%%%%%%%%%%%%%%%%%%%%%%%%%%%%%%%%%%%%%%%%%%%%%%%%%%%%%%%%%%%%%%%%

\clearpage
\section{Introduction}

E-commerce has exploded in recent times, achieving unbelievable global scale, delivery speed, and system complexity.
The short-term operations of a typical e-commerce giant involves
pulling inventory from suppliers into its fulfillment centers (FC's), including retail stores that can also be used to fulfill online orders;
awaiting purchases from online customers, which can be influenced by a powerful search/recommendation engine;
and finally delivering the goods to the customer's doorstep, through a flexible transportation system that allows different FC's in the network to be used for fulfilling demand from any particular region.

This paper focuses on the final part of these operations,
which is the problem of dynamically dispatching incoming customer orders to FC's,
while treating this customer demand (as influenced by search/recommendation) and inventory replenishment as exogenous.
The decision is on how to dynamically allocate finite inventories, of multiple items each of which has been placed in multiple FC's, over a finite time horizon, representing the duration until the next inventory replenishment.
The objective is to minimize the total costs from fulfillment and inventory stockouts.

This dynamic fulfillment problem is challenging for several reasons.
First, decisions must be made with consideration of the future orders to come, since depleting inventories at the wrong places can set off a chain reaction of long-distance and split shipments, as originally demonstrated by \citet{xu2009benefits}.
Moreover, due to the uncertainty in future orders, forward-lookingness requires a high-dimensional stochastic dynamic program that is intractable to solve, as noted by \citet{acimovic2019fulfillment}.
%Meanwhile, even a myopic strategy like using the minimum number of FC's to satisfy each incoming order, without consideration of future orders, can be computationally hard.
Finally, the mere scale and speed of the problem restricts us to fast and simple heuristics, with more elaborate optimizations exacerbating the issue of system complexity.

In light of these challenges, a prevailing approach to the dynamic fulfillment problem is deterministic-relaxation-based, as pioneered by \citet{jasin2015lp}.
Namely, a linear program (LP) that views the system as deterministic is written, describing inventory levels of every item at every FC, and expected demands at different regions which
%could vary over time and
includes information about items frequently purchased together in the same order.
The objective captures fixed shipping costs (mostly dependent on the number of distinct FC's used to fulfill an order), variable shipping costs (dependent on items and distances), and shortage costs (dependent on penalties paid for orders not fulfilled).
The LP is then solved, providing a ``master plan'' of matching supply to demand, which prescribes
for different orders from different regions, how frequently each FC should be used to fulfill each item in that order.
As orders come in real-time, \citet{jasin2015lp} randomly dispatch the items to FC's, making sure to follow the fulfillment frequencies outlined in the LP's plan.

Although seemingly uninformed, this \textit{randomized fulfillment} approach is simple, fast, and highly parallelizable as it does not require real-time inventory information across the network once the LP solution is given.
Under large system scales, it also pays variable shipping and shortage costs similar to what is outlined in the LP.
However, fixed costs remain a challenge---the problem of covering all the items in an order using a small number of distinct FC's was already difficult, and the LP's fulfillment frequencies now impose additional constraints.
Moreover, it has been shown that fixed costs, capturing the number of boxes from different origins shipped, constitute the majority of e-commerce fulfillment costs \citep{xu2009benefits,jasin2015lp}, so this presents a major issue.
The seminal insight, due to \citet{jasin2015lp}, is that these frequencies are actually helpful---when using them to randomly assign an FC to each item, if \textit{positive correlation} is induced in the assignments across items, then many items end up assigned to the same FC and not many distinct FC's are used.
The authors derive an intricate method for inducing this correlation.

Despite its significance and impact on subsequent work \citep[e.g.][]{lei2018joint,lei2021joint,zhao2020multi}, to the best of our knowledge, the correlation method of \citet{jasin2015lp} has never been substantially improved, until now.
This paper derives a new correlation method
that is intuitively simpler, computationally faster, and achieves tight performance in two different regimes.

\subsection{Correlated Rounding Problem of \citet{jasin2015lp}} \label{sec:introJasin}

Consider a single order (from a particular region at a particular time) consisting of $q$ items.
For each item in the order, denoted using $i\in\{1,\ldots,q\}$, we are told the fraction of time $u_{ki}$ that it must be fulfilled from each FC $k\in\{1,\ldots,K\}$.
Every item must be fulfilled, so $\sum_k u_{ki}=1$ for all $i$.
We must randomly choose an FC for each item $i$ according to these probabilities $u_{ki}$, and
an FC is \textit{used} if any item is assigned to it (meaning we would ship a box out of that FC).

Intuitively, the goal is to not use many distinct FC's.
This is formalized as no FC being used with a probability greater than necessary.
Specifically, for each FC $k$, it has to used with probability at least $u_{ki}$ to fulfill any item $i$, and hence $y_k:=\max_i u_{ki}$ is a lower bound on its probability of being used.
A method that randomly assigns every item $i$ to an FC following its marginal probability vector $(u_{ki})_{k=1}^K$ is called a \textit{rounding} scheme, and the rounding scheme is said to be \textit{$\alpha$-competitive} if it uses every FC $k$ with probability at most $\alpha\cdot y_k$, for some $\alpha\ge1$.  Here, $\alpha$ is referred to as the \textit{guarantee} of the rounding scheme, which is ideally as small as possible.

A naive rounding scheme is to \textit{independently} draw an FC for each item.
However, such a random outcome likely uses many distinct FC's, causing FC's to be used more frequently than necessary, and \citet{jasin2015lp} show that the guarantee of the independent rounding scheme can be as bad as $\alpha=q$ on a $q$-item order.
\citet{jasin2015lp} derive an improved rounding scheme that \textit{correlates} positively the FC's drawn across items, so that the random outcome uses fewer distinct FC's.  They establish that given any $q$ marginal distributions over FC's, this correlated rounding scheme is $\approx q/4$-competitive, improving the guarantee of the naive rounding by a factor of 4.

In this paper we derive two new correlated rounding schemes.
The first is $(1+\ln(q))$-competitive, completely improving the earlier guarantees in terms of order-dependence on $q$---from linear to logarithmic.
The second is $d$-competitive, where $d$ is a \textit{sparsity} parameter that describes the maximum number of options that any item has in terms of where to be fulfilled, i.e. $d=\max_i|\{k:u_{ki}>0\}|$.
Both of these guarantees are tight
for the correlated rounding problem,
as we will show.

\subsubsection*{Implications for dynamic fulfillment.}
Our rounding schemes are directly applicable to the original dynamic fulfillment problem via the approach of \citet{jasin2015lp}.
Indeed, each incoming order can be separately and randomly dispatched, using our choice of rounding scheme.
The results of \citet{jasin2015lp} then imply that in the dynamic fulfillment problem, the total cost paid is asymptotically at most $\beta$ times the optimum, where $\beta$ is a constant that depends on the average value of $\min\{1+\ln(q),d\}$ across orders (different orders have different sizes $q$ and sparsity parameters $d$, and we can choose the rounding scheme with the better guarantee between $1+\ln(q)$ and $d$ for each incoming order).
As a special case, if the largest order has size $\overline{q}$, then the guarantee is $1+\ln(\overline{q})$, matching computational hardness results for the dynamic fulfillment problem even when there is a single order.
Further details can be found in \textbf{\Cref{sec:multiOrder}}.

We note, however, that reducing everything down to the correlated rounding subroutine is not the only approach to dynamic fulfillment.  Indeed, the correlated rounding problem imposes frequency constraints on every order (that every item $i$ in every order is assigned to each FC $k$ with marginal probability exactly $u_{ki}$), with the rationale being that the variable shipping and inventory shortage costs become relatively inconsequential under large system scales;
unfortunately, this can be restrictive compared to some alternative approaches, as summarized in \citet{acimovic2019fulfillment}.
% may induce more split shipments and hence higher fixed shipping costs than necessary.
Nonetheless, this simple and fast approach performs well numerically in realistic setups, as shown in \citet{jasin2015lp}.
In \textbf{\Cref{sec:numExp}}, we show using the same setups that our new rounding schemes robustly bolster the performance of the randomized fulfillment approach, while shortening code and runtimes.
Our code is made publicly available.

\subsection{Main Idea behind New Rounding Schemes and Analysis} \label{sec:intuit}

Another benefit of our rounding schemes is that they have a simple intuition---each FC draws a random ``opening time'', and each item is assigned to the first FC that it sees open under its own, item-specific ``time dilation''.
%By contrast, the correlation method of \citet{jasin2015lp} based on line partitions, while clever and beautiful, is to our understanding not simple.
We now describe in detail our two rounding schemes and analysis.

Recall that we are trying to induce positive correlation in the FC's assigned across items.
To do this, we imagine a process where
each FC is initially closed, and opens at a random time.
Items are assigned to the first FC that they see open.
Importantly, each item $i$ views the openings of FC's on its own \textit{dilated} time scale, calibrated so that the probability of it seeing any FC $k$ open first is exactly $u_{ki}$.
Because an FC opening early means that it will be seen first by more (but not necessarily all) items, this induces positive correlation in the FC's assigned across different items.

To make this precise, for each FC $k$, we draw its opening time $E_k$ independently from an Exponential distribution with mean $1/y_k$, where $y_k:=\max_i u_{ki}$.
We then define the dilated time scale for an item $i$ as:
it sees each FC $k$ open at time $\frac{y_k}{u_{ki}}E_k$, which we note is no earlier than $E_k$, since $\frac{y_k}{u_{ki}}\ge1$.  (If $u_{ki}=0$, then $\frac{y_k}{u_{ki}}E_k=\infty$, and item $i$ never sees FC $k$ open.)
The dilated opening times $\frac{y_k}{u_{ki}}E_k$ are Exponentially distributed with means $\frac{y_k}{u_{ki}}\cdot\frac{1}{y_k}=\frac{1}{u_{ki}}$, and independent across $k$.
Through the lens of Poisson processes, it is easy to see that the probability of each FC $k$ arriving first into the view of item $i$ is exactly $\frac{u_{ki}}{u_{1i}+\cdots+u_{Ki}}=u_{ki}$, as desired.

The Poisson lens also helps us upper-bound the probability of an FC $k$ getting used at all.
Indeed, since an FC $k$ can only be seen at times \textit{later} than $E_k$, it can only get used if it arrives when at least one item is still waiting, an event whose probability is exponentially decaying over time.
Unfortunately, random variable $E_k$ is correlated with the latter event, making the analysis complicated. 
To fix this, we instead consider a related process where FC $k$ is ``repeatedly opening'' following a Poisson process of rate $y_k$, which allows us to exploit the memoryless property and take an elementary integral to show that the probability of FC $k$ opening is at most $(1+\ln(q))y_k$, completing our sketch of why our first rounding scheme is $(1+\ln(q))$-competitive.

To motivate our second rounding scheme, we note that the preceding analysis is poor when $q$ is enormous, because for a long time at least one item will still be waiting, during which FC openings will result in usage.
Therefore, we consider a modified scheme where each FC $k$ is ``forced open'' at time $1/y_k$, even if $E_k>1/y_k$.
For each item $i$, it will see each FC $k$ forced open at time $\frac{y_k}{u_{ki}}\cdot\frac{1}{y_k}=\frac{1}{u_{ki}}$.
Therefore, item $i$ will get ``force-assigned'' by time $\frac{1}{\max_ku_{ki}}$, and all items will be force-assigned by time $\alpha:=\frac{1}{\min_i\max_ku_{ki}}$, regardless of how many items there are.
Moreover, if $d$ is an upper bound on $|\{k:u_{ki}>0\}|$, then $\max_ku_{ki}\ge 1/d$ for all $i$, and hence $\alpha\le d$.
The fact that all items are assigned by time $d$ w.p.~1 allows us to show that no FC gets used with probability more than $dy_k$.

However, these forced openings cause each item $i$ to be over-fulfilled from the FC $m(i)$ that it would first see forced open.
Therefore, we make a second modification where for each item $i$, if the over-fulfilled FC $m(i)$ were to ``naturally'' open (i.e. $E_{m(i)}<1/y_{m(i)}$), then it is \textit{hidden} from the view of item $i$ (until it is forced open) with some likelihood.
This likelihood can be calibrated so that $i$ ends up seeing every FC $k$ open first with probability exactly $u_{ki}$, as desired.

\subsection{Further Technical Details and Relationship with Set Cover} \label{sec:allResults}

We now outline all our new results for the correlated rounding problem and the related technical results.
\begin{itemize}
\item Our main results are a $(1+\ln(q))$-competitive rounding scheme, and a $d$-competitive rounding scheme (where $d$ denotes the sparsity parameter $\max_i|\{k:u_{ki}>0\}|$).  These rounding schemes and their analyses are presented in \textbf{\Cref{sec:schemes}}.
\item The exact guarantee for the rounding scheme of \citet{jasin2015lp} is given by a function $B$ of the order size, where $B(q)=\frac{(q+1)^2}{4q}$ if $q$ is odd and $B(q)=\frac{q+2}{4}$ if $q$ is even.
For small values of $q$, this is better than our guarantee of $1+\ln(q)$; e.g.\ if $q=2$ then $B(1)=1$.
\item Both of our rounding schemes have a runtime of $O(qK)$.  By contrast, the rounding scheme of \citet{jasin2015lp} has a runtime of $O(q^2 K)$, containing a loop that is quadratic in the number of items $q$.
\item If there are only two FC's, i.e.\ $K=2$, then a 1-competitive rounding scheme was recently discovered by \citet{zhao2020multi}.
In this scenario, our second rounding scheme would only be 2-competitive, since $d=K=2$.  However, we emphasize that parameter $d$ represents the maximum number of distinct FC's that hold an item and can generally be much smaller than $K$, whereas their rounding scheme only works when $K=2$.
\item In \textbf{\Cref{sec:exactExponentialSoln}}, we establish an additional result that computes the optimal guarantee $\alpha$ and rounding scheme for a given instance, using an LP of size $O(2^K)$.  \citet{jasin2015lp} also show how to compute instance-optimal schemes, using an LP of size $O(K^q)$.  While both are exponentially-sized, our LP can be applied when $K$ is small; theirs can be applied when $q$ is small.
\end{itemize}

\subsubsection*{Relating the correlated rounding problem to Set Cover.}
%This allows us to establish the tightness of our guarantees, and also uncover a new randomized rounding algorithm for Set Cover.
\begin{itemize}
\item In \textbf{\Cref{sec:setCover}}, we show that an $\alpha$-competitive rounding scheme implies a procedure for rounding a fractional Set Cover solution into a randomized cover, that is feasible w.p.~1, and has no set chosen with probability more than $\alpha$ times its fractional weight.
\item Therefore, we can leverage hardness results from Set Cover to show that an $\alpha$-competitive rounding scheme must have $\alpha=\Omega(\log (q))$ and $\alpha\ge d$.
The former lower bound establishes our $(1+\ln(q))$-competitive rounding scheme to be order-optimal in $q$, while the latter lower bound establishes our $d$-competitive rounding scheme to be exactly tight in $d$.
\item Our $(1+\ln(q))$-competitive rounding scheme also improves guarantees in the aforementioned randomized rounding problem for Set Cover.
To the best of our knowledge, existing rounding methods for Set Cover take each set with probability at least $\ln(q)+\omega(1)$ times its fractional weight \citep{raghavan1987randomized}; see also \citet{motwani1995randomized} and \citet[Sec~14.2]{vazirani2001approximation}.
Although our improvement to $\ln(q)+1$ is only in lower-order terms, our approach via the correlated rounding problem is both new, and simpler than many of the commonly-taught methods.
\end{itemize}

We note that for the Set Cover problem itself, which has nothing to do with randomization, the Greedy algorithm has a guarantee of $1+1/2+\cdots+1/q$, which is slightly smaller (better) than our $1+\ln(q)$.
%A fractional Set Cover solution is also easily converted into an integral one while losing a factor of at most $d$.
Nonetheless, we believe these connections highlight how the correlated rounding problem is a harder version of Set Cover---in which a randomized solution, that must satisfy constraints on how often each set is used to cover each element, is required.
Furthermore, it is interesting to us that a modern problem from e-commerce practice, identified by \citet{jasin2015lp}, can lead us to improve randomized rounding schemes for the age-old Set Cover problem from CS theory.

\subsection{Further Related Work} \label{sec:relatedWork}

The dynamic fulfillment problem, and in particular the correlated rounding approach, is more challenging and relevant in large fulfillment networks.
Fulfillment networks have been getting larger with the advent of omni-channel retailing, which allows for online orders to be fulfilled from small retail stores \citep{acimovic2019fulfillment}.
Although order sizes have been decreasing with the advent of fast shipping, online retailers have been making greater efforts to \textit{delay} fulfillment and \textit{consolidate} multiple orders into one before fulfilling \citep{wei2021shipping,Wang22}.
Consequently, the dynamic fulfillment problem with multi-item orders and \textit{flexibility} in how to fulfill them is as relevant as ever \citep{devalve2021understanding}.

In terms of the overall LP-based approach that justifies the correlated rounding problem, we should note that LP-based approaches are also heavily employed in the revenue management literature \citep[see e.g.][]{talluri2004theory}.  They enjoy many benefits such as scalability and ability to incorporate side constraints, and the given probabilities $u_{ki}$ can always be updated over time through re-solving \citep[see e.g.][]{jasin2012re} to adjust for updated inventories and demand predictions over time.
An early work advocating for the LP-based approach in e-commerce fulfillment is \citet{acimovic2015making}.
Very recently, \citet{amil2022multi} propose a novel LP that can be used in place of the standard one, which we discuss at the end of \Cref{sec:multiOrder}.

\section{Formal Specification and Analysis of Rounding Schemes} \label{sec:schemes}

We recap the correlated rounding problem from the Introduction, our main object of study.

\begin{definition}[Recap of Problem, Notation, and Terminology]\
\begin{itemize}
\item An \textit{instance} of the $\alpha$-competitive rounding scheme problem consists of $q$ marginal distributions over $K$ FC's, given by probabilities $u_{ki}$ satisfying $\sum_{k=1}^Ku_{ki}=1$ for all $i=1,\ldots,q$.
\item A \textit{rounding scheme} must randomly assign each item $i$ to an FC $Z_i\in\{1,\ldots,K\}$, satisfying the marginal conditions $\Pr[Z_i=k]=u_{ki}$ for all $i$ and $k$.
\item An FC $k$ is \textit{used} if any item is assigned to it, denoted by the event $\bigcup_{i=1,\ldots,q}(Z_i=k)$, which must occur with probability at least $y_k:=\max_i u_{ki}$.  Assume without loss that $y_k>0$ for all $k$.
\item A rounding scheme is \textit{$\alpha$-competitive} if given any instance, it uses each FC $k$ with probability at most $\alpha\cdot y_k$.  The guarantee $\alpha$ can depend on parameters of the instance.
\item The \textit{sparsity} parameter of an instance is defined as $d=\max_i|\{k:u_{ki}>0\}|$, the maximum number of distinct FC's that one item $i$ could get assigned to.
\end{itemize}
\end{definition}

We now provide efficient algorithmic specifications of our rounding schemes and analyze them.
We believe both our algorithms and proofs to be quite intuitive, and will frequently provide proof sketches that refer back to the intuition from \Cref{sec:intuit}, where items are waiting for FC's to open on their own dilated time scales.

\subsection{$(1+\ln(q))$-competitive Rounding Scheme} \label{sec:log_n}

Our rounding scheme is specified in \Cref{alg:1}.  Relating back to the intuitive description, $E_k$ is the time at which FC $k$ opens, and $\frac{y_k}{u_{ki}}E_k$ is the \textit{delayed} time (since $\frac{y_k}{u_{ki}}\ge1$) at which item $i$ sees it open, with $\frac{y_k}{u_{ki}}E_k=\infty$ if $u_{ki}=0$.
Every item is assigned to the first FC that it sees open.

\begin{algorithm}
\caption{$(1+\ln(q))$-competitive Rounding Scheme} \label{alg:1}
\begin{algorithmic}
\For{$k=1,\ldots,K$}
	\State $E_k\leftarrow$ independent draw from Exponential distribution with mean $1/y_k$
\EndFor
\For{$i=1,\ldots,q$}
	\State $Z_i\leftarrow\argmin_{k=1,\ldots,K}\frac{y_k}{u_{ki}}E_k$ \Comment{Break ties arbitrarily}
\EndFor
\end{algorithmic}
\end{algorithm}

We now prove that \Cref{alg:1} is a $(1+\ln(q))$-competitive Rounding Scheme, where $q$ is the number of items.
To establish the marginals condition, we use the interpretation that from the perspective of any individual item, the FC's open according to independent Poisson processes.

\begin{lemma} \label{lem:log_nFeas}
Under \Cref{alg:1}, $\Pr[Z_i=k]=u_{ki}$ for all $i=1,\ldots,q$ and $k=1,\ldots,K$.
\end{lemma}

\proof{Proof of \Cref{lem:log_nFeas}.}
Consider the perspective of any item $i$.  Index $Z_i$ is determined by the smallest realization among $\{\frac{y_k}{u_{ki}}E_k:k=1,\ldots,K\}$, which are independent Exponential random variables with means $\{\frac{1}{u_{ki}}:k=1,\ldots,K\}$.
Equivalently, $Z_i$ is determined by the first arrival among independent Poisson processes with rates $\{u_{ki}:k=1,\ldots,K\}$.
By the Poisson merging theorem, each Poisson process $k$ will be the first to arrive with probability $\frac{u_{ki}}{u_{1i}+\cdots+u_{Ki}}$, which equals $u_{ki}$ since $u_{1i}+\cdots+u_{Ki}=1$.  Therefore, $\Pr[Z_i=k]=u_{ki}$ for all $k=1,\ldots,K$, completing the proof.
\Halmos\endproof

We now prove an intermediate lemma that, intuitively, bounds the probability of any item $i$ still ``waiting'' (to be assigned to an FC) up to time $t$, which can be expressed as the event $(\min_k\frac{y_k}{u_{ki}}E_k\ge t)$.
The final statement then takes a union bound of having any item still waiting, which intuitively is not too loose since these events are positively correlated---one item waiting implies that FC's were late to open, which makes other items more likely to also be waiting.

\begin{lemma} \label{lem:log_nTime}
Under \Cref{alg:1}, $\Pr[\bigcup_{i=1}^q(\min_k\frac{y_k}{u_{ki}}E_k\ge t)]\le qe^{-t}$ for all $t\ge0$.
\end{lemma}

\proof{Proof of \Cref{lem:log_nTime}.}
First consider any item $i$.  Random variables $\{\frac{y_k}{u_{ki}}E_k:k=1,\ldots,K\}$ are independent and Exponentially distributed with means $\{\frac{1}{u_{ki}}:k=1,\ldots,K\}$.
Therefore, $\min_k\frac{y_k}{u_{ki}}E_k$ is Exponentially distributed with mean $\frac{1}{u_{1i}+\cdots+u_{Ki}}=1$.
Consequently, $\Pr[\min_k\frac{y_k}{u_{ki}}E_k\ge t]=e^{-t}$, and by the union bound, $\Pr[\bigcup_{i=1}^q(\min_k\frac{y_k}{u_{ki}}E_k\ge t)]\le qe^{-t}$, completing the proof.
\Halmos\endproof

We are now ready to prove our main result for \Cref{alg:1}.  Although technical, the argument uses a simple intuitive trick.
\Cref{lem:log_nTime} has upper-bounded the probability of \textit{any} item still waiting at a time $t$.
If an FC $k$ opens at a time when no item is still waiting, then it is guaranteed to not get used (since items can only see it open at a \textit{delayed} time).
Unfortunately, the opening time of an FC $k$ is correlated with the event of having an item still waiting.
To fix this, we imagine FC $k$ as ``repeatedly opening'' following a Poisson process of rate $y_k$, with it being ``used'' every time it opens as long as there is an item still waiting.
Since Poisson processes are memoryless, this now de-correlates the events of FC $k$ opening from the event of still having an item waiting.
\Cref{lem:log_nTime} can then apply, and the analysis finishes by taking an integral.
The formal proof is presented below.

\begin{theorem} \label{thm:log_n}
\Cref{alg:1} is a $(1+\ln(q))$-competitive rounding scheme with runtime $O(qK)$.
\end{theorem}

\proof{Proof of \Cref{thm:log_n}.}
The runtime is $O(qK)$ because taking the $\argmin$ over $k=1,\ldots,K$ for all $i=1,\ldots,q$ is the bottleneck operation in \Cref{alg:1}.
Meanwhile, \Cref{lem:log_nFeas} has already shown that the marginals condition is satisfied.
It remains to show that $\Pr[\bigcup_{i=1,\ldots,q}(Z_i=k)] \le\alpha y_k$ for all $k$, with $\alpha=1+\ln(q)$.

Fix any FC $k$.  For all items $i$ with $u_{ki}>0$, event $Z_i=k$ can occur only if $k$ lies in the $\argmin$ in \Cref{alg:1}, i.e. if $\min_{k'}\frac{y_{k'}}{u_{k'i}}E_{k'}\ge\frac{y_k}{u_{ki}}E_k$.
We now rewrite this event as follows.
Define $S^1_k,S^2_k,\ldots$ to be the arrival times of a Poisson process of rate $y_k$.
More specifically, we will let $S^1_k=E_k$, and $S^{j+1}_k$ be the sum of $S^j_k$ with an independent Exponential random variable of mean $1/y_k$, for all $j\ge1$.
We can derive
\begin{align}
(Z_i=k) &\subseteq\left(\min_{k'}\frac{y_{k'}}{u_{k'i}}E_{k'}\ge\frac{y_k}{u_{ki}}E_k\right) \nonumber
\\ &=\left(\min_{k'\neq k}\frac{y_{k'}}{u_{k'i}}E_{k'}\ge\frac{y_k}{u_{ki}}S^1_k\right) \nonumber
\\ &=\bigcup_{j=1}^{\infty}\left(\min\left\{\min_{k'\neq k}\frac{y_{k'}}{u_{k'i}}E_{k'},\min_{j'<j}\frac{y_k}{u_{ki}}S^{j'}_k\right\}\ge\frac{y_k}{u_{ki}}S^j_k\right) \label{eqn:2552}
\end{align}
where the final equality~\eqref{eqn:2552} holds because the events with $j>1$ never occur (in particular, $\min_{j'<j}\frac{y_k}{u_{ki}}S^{j'}_k\ge\frac{y_k}{u_{ki}}S^j_k$ is impossible since $S^{j'}_k<S^j_k$).
The purpose of this vacuous decomposition is to later relax the event (by decreasing the RHS) and then apply the memorylessness property of Poisson processes.

We now take a union bound of events~\eqref{eqn:2552} over $i$, and analyze the probability of this union by conditioning on the event that $S^j_k=t$ for any $j\ge1$, over all times $t\ge0$.
Formally:
\begin{align*}
&\Pr\left[\bigcup_{i:u_{ki}>0}\bigcup_{j=1}^{\infty}\left(\min\left\{\min_{k'\neq k}\frac{y_{k'}}{u_{k'i}}E_{k'},\min_{j'<j}\frac{y_k}{u_{ki}}S^{j'}_k\right\}\ge\frac{y_k}{u_{ki}}S^j_k\right)\right]
\\ &=\int_0^{\infty}\Pr\left[\bigcup_{i:u_{ki}>0}\left(\min\left\{\min_{k'\neq k}\frac{y_{k'}}{u_{k'i}}E_{k'},\min_{j'<j}\frac{y_k}{u_{ki}}S^{j'}_k\right\}\ge\frac{y_k}{u_{ki}}t\right)\Bigg|\exists j:S^j_k=t\right]y_kdt
\\ &\le \int_0^{\infty}\Pr\left[\bigcup_{i:u_{ki}>0}\left(\min\left\{\min_{k'\neq k}\frac{y_{k'}}{u_{k'i}}E_{k'},\min_{j'<j}\frac{y_k}{u_{ki}}S^{j'}_k\right\}\ge t\right)\Bigg|\exists j:S^j_k=t\right]y_kdt
\\ &=\int_0^{\infty}\Pr\left[\bigcup_{i:u_{ki}>0}\min_{k'=1,\ldots,K}\frac{y_{k'}}{u_{k'i}}E_{k'}\ge t\right]y_kdt
\\ &\le y_k\int_0^{\infty}\min\{q e^{-t},1\}dt
\end{align*}
where the first equality holds because the PDF of the event $(\exists j:S^j_k=t)$ takes value $y_k$ for all $t$,
the first inequality holds because $\frac{y_k}{u_{ki}}\ge1$, the second equality applies the memorylessness property of Poisson processes, and the final inequality applies \Cref{lem:log_nTime} (along with the trivial upper bound of 1).
Note that this analysis holds for any FC $k=1,\ldots,K$.  Therefore, the proof is now completed by taking an elementary integral:
\begin{align*}
\int_0^{\infty}\min\{q e^{-t},1\}dt
 &=\ln(q)+\int_{\ln(q)}^{\infty}q e^{-t}dt
\\ &=\ln(q)+q e^{-\ln(q)}
\\ &=1+\ln(q).
\end{align*}
\Halmos\endproof

\begin{remark}
Our \Cref{alg:1} and \Cref{thm:log_n} close the gap that was left open by the correlated rounding scheme of \citet{jasin2015lp}, whose guarantee grew linearly (instead of logarithmically) in the number of items $q$.
Their scheme partitions the [0,1] interval and makes the positive correlation in the FC's assigned very explicit.
By contrast, our rounding schemes are based on a ``trick'' of dilating memoryless random variables, and the positive correlation is implicit.
Our trick is designed to facilitate a short analysis, which closes the gap in the correlated rounding problem.
\end{remark}

\subsection{$d$-competitive Rounding Scheme} \label{sec:d}

Our modified rounding scheme is specified in \Cref{alg:2}.
Relating back to the intuitive description from \Cref{sec:intuit}, $m(i)$ is the first FC that item $i$ would see ``forced'' open, which it would get assigned to if it was still unassigned at that point.
$X_{ki}$ is a random variable denoting the time at which item $i$ sees FC $k$ open, which equals $\frac{y_k}{u_{ki}}E_k$ like before if $k\neq m(i)$.
On the other hand, $X_{m(i),i}$ is upper-bounded by $1/u_{m(i),i}$, as that is when item $i$ would see FC $m(i)$ forced open.
The final wrinkle is that if FC $m(i)$ were to ``naturally'' open before it is forced open, then it needs to be \textit{hidden} from $i$'s view (until it is forced open) with some probability, which is indicated by the random variable $H_i$.
Finally, every item is assigned to the first FC that it sees open, after taking into consideration hiding and forced opening.

\begin{algorithm}
\caption{$d$-competitive Rounding Scheme} \label{alg:2}
\begin{algorithmic}
\For{$k=1,\ldots,K$}
	\State $E_k\leftarrow$ independent draw from Exponential distribution with mean $1/y_k$
\EndFor
\For{$i=1,\ldots,q$}
	\State $m(i)\leftarrow\argmax_k u_{ki}$
	\For{$k=1,\ldots,K,\ k\neq m(i)$}
		\State $X_{ki}\leftarrow\frac{y_k}{u_{ki}}E_k$
	\EndFor
	\State $H_i\leftarrow$ independent draw from Bernoulli distribution with mean $\frac{1-u_{m(i),i}}{1-u_{m(i),i}+u_{m(i),i}e^{1/u_{m(i),i}}-e}$

	\Comment{$H_i=1$ means FC $m(i)$ is hidden from item $i$ until the FC is forced open at time $1/y_{m(i)}$}
	\State $X_{m(i),i}\leftarrow\frac{y_{m(i)}}{u_{m(i),i}}\min\{\frac{E_{m(i)}}{1-H_i},\frac{1}{y_{m(i)}}\}$ \Comment{$H_i=1$ means $\frac{E_{m(i)}}{1-H_i}=\infty$, and hence $X_{m(i),i}=\frac{1}{u_{m(i),i}}$}
	\State $Z_i\leftarrow\argmin_{k=1,\ldots,K} X_{ki}$
\EndFor
\end{algorithmic}
\end{algorithm}

It can be checked that the probability with which $H_i=1$ defined in \Cref{alg:2} does indeed lie in $[0,1]$ for all possible values of $u_{m(i),i}\in(0,1]$.
The hiding probability is in fact increasing in $u_{m(i),i}$, which is intuitive because a larger value of $u_{m(i),i}$ implies an earlier forced opening, suggesting that FC $m(i)$ should be hidden more often to prevent it from over-fulfilling item $i$.
We now prove that this hiding probability has been calibrated so that the marginals condition is satisfied exactly.

\begin{lemma} \label{lem:dFeas}
Under \Cref{alg:2}, $\Pr[Z_i=k]=u_{ki}$ for all $i=1,\ldots,q$ and $k=1,\ldots,K$.
\end{lemma}

\proof{Proof of \Cref{lem:dFeas}.}
Fix any item $i$.  We show that $\Pr[Z_i=k]=u_{ki}$ for all $k\neq m(i)$, which would automatically imply $\Pr[Z_i=m(i)]=1-\sum_{k\neq m(i)}\Pr[Z_i=k]=1-\sum_{k\neq m(i)}u_{ki}=u_{m(i),i}$.  We need to consider two cases: $H_i=1$ and $H_i=0$.  Hereafter omit index $i$.

First, if $H=1$, then the item does not observe FC $m$ before time $1/u_m$.
Therefore, $Z=k$ if and only if $X_k$ is the smallest among random variables $\{X_{k'}:k'\neq m\}$ and also $X_k<1/u_m$.
Recall that $X_{k'}$ is Exponentially distributed with mean $1/u_{k'}$ for all $k'\neq m$, and the $X_{k'}$'s are independent across $k'$.
Therefore, the probability that $\min_{k'\neq m}X_{k'}<1/u_m$ is equal to the probability that a Poisson process with rate $\sum_{k'\neq m}u_{k'}=1-u_m$ generates an arrival before time $1/u_m$, which occurs w.p.~$1-e^{-(1-u_m)/u_m}$.
Conditional on this, the probability that $\min_{k'\neq m}X_{k'}=X_k$ is exactly $\frac{u_k}{1-u_m}$, by the Poisson merging theorem.
Therefore,
\begin{align} \label{eqn:8902}
\Pr[Z=k|H=1]=(1-e^{-(1-u_m)/u_m})\frac{u_k}{1-u_m}.
\end{align}

Otherwise, if $H=0$, then the item observes all FC's before time $1/u_m$.
In this case, $Z=k$ if and only if $X_k$ is the smallest among all random variables $\{X_{k'}:k'=1,\ldots,K\}$ and also $X_k<1/u_m$.
By a similar argument as above, the probability that $\min_{k'=1,\ldots,K}X_{k'}<1/u_m$ is $1-e^{1/u_m}$, and conditional on this, the probability that $\min_{k'=1,\ldots,K}X_{k'}=X_k$ is $u_k$.
Therefore,
\begin{align} \label{eqn:8903}
\Pr[Z=k|H=0]=(1-e^{1/u_m})u_k.
\end{align}

Let $\eta$ denote $\frac{1-u_{m}}{1-u_{m}+u_{m}e^{1/u_{m}}-e}$, the probability that $H=1$.  Combining~\eqref{eqn:8902} and~\eqref{eqn:8903}, we derive
\begin{align*}
\Pr[Z_i=k]
&=\eta(1-e^{-(1-u_m)/u_m})\frac{u_k}{1-u_m}+(1-\eta)(1-e^{-1/u_m})u_k
\\ &=u_k\left(1-e^{-1/u_m}+\eta\left(\frac{1-e^{-(1-u_m)/u_m}}{1-u_m}-(1-e^{-1/u_m})\right)\right)
\\ &=u_k\left(1-e^{-1/u_m}+\eta\cdot\frac{-e^{-(1-u_m)/u_m}+u_m+e^{-1/u_m}-u_me^{-1/u_m}}{1-u_m}\right)
\\ &=u_k\left(1-e^{-1/u_m}+e^{-1/u_m}\eta\cdot\frac{1-u_m+u_me^{1/u_m}-e}{1-u_m}\right)
\\ &=u_k
\end{align*}
which completes the proof.
\Halmos\endproof

We now prove our main result for \Cref{alg:2}.
We establish the stronger guarantee of $\alpha=\frac{1}{\min_i\max_k u_{ki}}$, which is easily seen to be at most $d$ since $\max_k u_{ki}\ge1/d$ for all $i$.
The proof sketch is that due to the forced openings, all items are guaranteed to be assigned by time $\alpha$.
Therefore, an FC $k$ can only get used is it opens before time $\alpha$ (since items can only see it open with a delay), which occurs with probability no greater than $\alpha y_k$.

\begin{theorem} \label{thm:d}
\Cref{alg:2} is an $\frac{1}{\min_i\max_k u_{ki}}$-competitive rounding scheme with runtime $O(qK)$.
\end{theorem}

\proof{Proof of \Cref{thm:d}.}
The runtime is $O(qK)$, because inside the loop for $i=1,\ldots,q$ in \Cref{alg:2}, there are three bottleneck operations that each take time $O(K)$: the defining of $m(i)$, the inner loop for $k$, and the defining of $Z_i$.
Meanwhile, \Cref{lem:dFeas} has already shown that the marginals condition is satisfied.
It remains to show that $\Pr[\bigcup_{i=1,\ldots,q}(Z_i=k)] \le\alpha y_k$ for all $k$, with $\alpha=\frac{1}{\min_i\max_{k'} u_{k'i}}$.

Fix an FC $k$.
If $y_k\ge\min_i\max_{k'} u_{k'i}$, then $\alpha y_k\ge1$ and there is nothing to prove.
Therefore, assume $y_k<\min_i\max_{k'} u_{k'i}$, and we must show that $\Pr[\bigcup_{i=1,\ldots,q}(Z_i=k)] \le\alpha y_k$.
Since $y_k<\max_{k'} u_{k'i}$ for all $i$, we know that $k\neq m(i)$ for all $i$.
Thus, we have $X_{ki}=\frac{y_k}{u_{ki}}E_k$ for all $i$, and can write
\begin{align*}
(Z_i=k)
&\subseteq(\frac{y_k}{u_{ki}}E_k\le\min_{k'=1,\ldots,K}X_{k'i})
\\ &\subseteq(\frac{y_k}{u_{ki}}E_k\le X_{m(i),i})
\\ &\subseteq(\frac{y_k}{u_{ki}}E_k\le 1/u_{m(i),i})
\\ &=(\frac{y_k}{u_{ki}}E_k\le\frac{1}{\max_{k'}u_{k'i}})
\\ &\subseteq(\frac{y_k}{u_{ki}}E_k\le\alpha)
\\ &\subseteq(E_k\le\alpha)
\end{align*}
with the final relationship between events holding because $\frac{y_k}{u_{ki}}\ge1$.
Note that the final event is independent of $i$.
Therefore,
\begin{align*}
\Pr\left[\bigcup_{i=1,\ldots,q}(Z_i=k)\right]
\le\Pr[E_k\le\alpha]=1-e^{-\alpha y_k}
\end{align*}
which is at most $\alpha y_k$, completing the proof.
\Halmos\endproof

\section{Connections with Set Cover} \label{sec:setCover}

In this \namecref{sec:setCover} we establish our rounding schemes to be order-optimal in terms of the dependence on $q$ or $d$, by reducing our problem to that of rounding a fractional solution for Set Cover.
We first define the Set Cover problem and some basic concepts using our language of items and FC's.
We refer to \citet{vazirani2001approximation} for further background.

\begin{problem}[Weighted Set Cover] \label{prob:setCover}
There are items $i=1,\ldots,q$ to be covered by FC's $k=1,\ldots,K$.
Each FC $k$ requires a fixed cost of $c_k$ to open, and if opened, can cover all items in a set $U_k\subseteq\{1,\ldots,q\}$.
The objective is to find a collection of FC's to open, that covers all the items, and minimizes the sum of fixed costs paid for opening FC's.
The \textit{sparsity} of the instance is defined as $d:=\max_i|\{k:i\in U_k\}|$, the maximum number of different FC's that an item $i$ can be covered by.
\end{problem}

\begin{definition}[Set Cover Linear/Integer Programs]
The following Integer Program is called the \textit{Set Cover IP}.  In it, binary variable $y_k$ represents FC $k$ being opened.  It is an equivalent formulation of the Weighted Set Cover problem.
\begin{align}
\min\ \sum_{k=1}^Kc_ky_k \nonumber
\\ \text{s.t. } \sum_{k:i\in U_k}y_k&\ge1 &\forall i=1,\ldots,q \label{sc:cover}
\\ y_k &\in\{0,1\} &\forall k=1,\ldots,K \label{sc:integral}
\end{align}
Meanwhile, the \textit{Set Cover LP} is defined as the relaxation of the Set Cover IP with constraint~\eqref{sc:integral} changed to $y_k\in[0,1]$, for all $K=1,\ldots,K$.
\end{definition}

We now define the problem of rounding a fractional solution for Set Cover, in a way that is analogous to an $\alpha$-competitive rounding scheme, except we will call it an $\alpha$-competitive ``covering'' scheme instead.

\begin{definition}[$\alpha$-competitive Covering Scheme] \label{def:coveringScheme}
For $\alpha\ge1$,
an \textit{$\alpha$-competitive covering scheme} is a method for constructing random variables $Y_1,\ldots,Y_K\in\{0,1\}$ satisfying
\begin{align}
\sum_{k:i\in U_k}Y_k&\ge1 &\forall i=1,\ldots,q,\ w.p.~1 \label{eqn:coveringSchemeFeasible}
\\ \bE[Y_k] &\le \alpha\cdot y_k &\forall k=1,\ldots,K \label{eqn:coveringSchemeOptimal}
\end{align}
given any feasible solution $(y_k)_{k=1}^K$ to the Set Cover LP.
\end{definition}

We now show that coming up with $\alpha$-competitive rounding schemes is a harder problem than
coming up with $\alpha$-competitive covering schemes.

\begin{lemma} \label{prop:reduction}
An $\alpha$-competitive rounding scheme can be efficiently applied as an $\alpha$-competitive covering scheme.
Moreover, any dependence of $\alpha$ on the parameters $q$ or $d$ translate over directly.
\end{lemma}

\proof{Proof of \Cref{prop:reduction}.}
Take any instance of Set Cover and a feasible solution $(y_k)_{k=1}^K$ to its LP.
For each item $i$, arbitrarily set $u_{ki}\in[0,y_k]$ for each FC $k$ that can cover it, so that $\sum_{k:i\in U_k} u_{ki}=1$.
We note that this is always possible since $y_k\ge0$ and $\sum_{k:i\in U_k} y_k\ge1$ by~\eqref{sc:cover}.
Meanwhile, set $u_{ki}=0$ if $i\notin U_k$.

The marginal distributions $(u_{k1})_{k=1}^K,\ldots,(u_{kn})_{k=1}^K$ now define an instance for an $\alpha$-competitive rounding scheme, with the same number of items $q$ and a sparsity $d$ that is no greater than before.
We apply the $\alpha$-competitive rounding scheme that is assumed to exist on this instance, and define random variables $Y_k=\bI(\bigcup_i(Z_i=k))$ for all $k=1,\ldots,K$.
By the definition of a rounding scheme, for each item $i$, we know that $Z_i=k$ is true for some index $k\in\{1,\ldots,K\}$, with $k\in U_k$ since otherwise $u_{ki}=0$.
Therefore, $Y_k=1$ for this index $k$ and condition~\eqref{eqn:coveringSchemeFeasible} for the covering scheme is satisfied.
Meanwhile, applying the definition of an $\alpha$-competitive rounding scheme, we have
$$
\bE[Y_k]=\Pr\left[\bigcup_i(Z_i=k)\right]\le\alpha\cdot\max_iu_{ki}\le y_k.
$$
We conclude that condition~\eqref{eqn:coveringSchemeOptimal} for the covering is satisfied.
We also note that if $\alpha$ depends on the sparsity parameter $d$, then the same guarantee continues to hold under the old sparsity parameter for Set Cover which is no less than $d$, completing the proof.
\Halmos\endproof

\subsection{Negative Results for $\alpha$-competitive Rounding Schemes} \label{sec:negSetCover}

Equipped with \Cref{prop:reduction}, we can now translate hardness results for the $\alpha$-competitive covering scheme problem into hardness results for the $\alpha$-competitive rounding scheme problem.

\begin{corollary}[of \Cref{prop:reduction}]
An $\alpha$-competitive covering scheme must have $\alpha=\Omega(\log (q))$ \citep[Ex.~13.4]{vazirani2001approximation}.
Therefore, an $\alpha$-competitive rounding scheme must also have $\alpha=\Omega(\log (q))$.
Consequently, the $(1+\ln(q))$-competitive rounding scheme established in \Cref{thm:log_n} achieves the order-optimal dependence on $q$.
\end{corollary}

\begin{proposition} \label{prop:setCover_d}
An $\alpha$-competitive covering scheme must have $\alpha\ge d$, where $d$ denotes the sparsity of the instance.
\end{proposition}

\proof{Proof of \Cref{prop:setCover_d}.}
Consider a Set Cover instance with $d$ fixed, $K$ large, and one item for each subset of $\{1,\ldots,K\}$ of size $d$.
Each such item can only be covered by the $d$ FC's in its corresponding subset, with the total number of items being $q=\binom{K}{d}$.
The sparsity of this instance is $d$ by definition.

Setting $y_k=1/d$ for all $k=1,\ldots,K$ forms a feasible solution to the Set Cover LP, since $|\{k:i\in U_k\}|=d$ for all items $i$, and hence LP constraints~\eqref{sc:cover} are satisfied.
On the other hand, any $\alpha$-competitive covering scheme must set $\sum_{k=1}^K Y_k>K-d$ w.p.~1, since otherwise there would be an uncovered item, violating~\eqref{eqn:coveringSchemeFeasible}.  Using the linearity of expectation, we derive
$$
K-d\le\sum_{k=1}^K\bE[Y_k]\le \sum_{k=1}^K\alpha\cdot y_k=K\alpha\frac{1}{d},
$$
with the second inequality coming from~\eqref{eqn:coveringSchemeOptimal}.
Therefore, $\alpha\ge d(1-\frac{d}{K}),$ with $\frac{d}{K}$ approaching for arbitrarily large $K$, completing the proof.
\Halmos\endproof

\begin{corollary}[of \Cref{prop:reduction} and \Cref{prop:setCover_d}]
An $\alpha$-competitive rounding scheme must have $\alpha\ge d$.
Consequently, the $d$-competitive rounding scheme established in \Cref{thm:d} achieves the optimal (not just order-optimal) dependence on $d$.
\end{corollary}

\section{Instance-Optimal Rounding Schemes} \label{sec:exactExponentialSoln}

The $(1+\ln(q))$-competitive and $d$-competitive rounding schemes discussed in \Cref{sec:schemes,sec:setCover} were only order-optimal in the worst case.
For a particular instance given by $q$ marginals over $\{1,\ldots,K\}$,
one could also consider the problem of computing the maximum guarantee $\alpha$ and rounding scheme that satisfies the marginal frequency constraints.

We formulate this problem using an LP with the following variables.
For all subsets $S$ of the FC's $\{1,\ldots,K\}$, let $z(S)$ denote the probability that exactly the set of FC's in $S$ get used.
For all $S\subseteq\{1,\ldots,K\}$, FC's $k\in S$, and items $i$, let $u_{ki}(S)$ denote the probability that the set of FC's in $S$ get used \textit{and} that item $i$ is fulfilled from FC $k\in S$.
The problem of minimizing $\alpha$ in an $\alpha$-competitive rounding scheme for this particular instance can then be formulated as
\begin{align}
\min\ \alpha \label{eqn:minObj}
\\ \text{s.t. }\sum_{k\in S}u_{ki}(S) &= z(S) &\forall S,i=1,\ldots,q \label{11}
\\ \sum_Su_{ki}(S) &=u_{ki} &k=1,\ldots,K,i=1,\ldots,q \label{22}
\\ \sum_{S\ni k}z(S) &\le \alpha \cdot y_k &\forall k=1,\ldots,K \label{29}
\\ \sum_Sz(S) &=1 \label{33}
\\ z(S) &\ge0 &\forall S \label{44}
\\ u_{ki}(S) &\ge0 &\forall S,k\in S, i=1,\ldots,q \label{55}
\end{align}
where constraints~\eqref{11} enforce that every item $i$ must be fulfilled from exactly one FC on each subset $S$, constraints~\eqref{22} and~\eqref{29} enforce the marginal and $\alpha$-competitive properties of a rounding scheme, constraints~\eqref{33}--\eqref{44} enforce that exactly one subset $S$ is selected, and last but not least,~\eqref{55} ensures that there is only a variable $u_{ki}(S)$ if $k\in S$.
%The LP does not enforce that every FC $k\in S$ actually gets used (i.e.\ has $u_{ki}(S)>0$ for some $i$), but note that if this is the case, then $k$ can be discarded from the set $S$ while decreasing fixed costs.

Our LP has size $O(nK2^K)$, which is exponential in $K$ but tractable if $K$ is a fixed constant.
\citet{jasin2015lp} derive an exponential-sized LP for the same purpose, except instead there is a variable for every possible \textit{mapping} from $\{1,\ldots,q\}$ to $\{1,\ldots,K\}$, for which there are $K^q$ possibilities.
Our LP's are more practical in situations where $K$ is small but $q$ is large, which is the case in the application of e.g.\ \citet{zhao2020multi}.

\section{$\alpha$-competitive Rounding Scheme applied to Dynamic Fulfillment} \label{sec:multiOrder}

In this \namecref{sec:multiOrder} we recap the general dynamic fulfillment problem from \citet{jasin2015lp}, and formalize the implication of our $\alpha$-competitive rounding schemes for the overall problem.

\subsubsection*{Problem definition.}
There is a horizon consisting of time steps $t=1,\ldots,T$, during which
items $i=1,\ldots,n$ are fulfilled from FC's $k=1,\ldots,K$.
Each item $i$ starts with $b_{ki}$ units of inventory at each FC $k$, with the end of the horizon representing the time at which inventories are replenished again.
Orders come from one of regions $j=1,\ldots,J$, and are described by a subset\footnote{
This section introduces the broader problem with $n$ items in the universe and orders $a$ which are subsets of $\{1,\ldots,n\}$.
The earlier \Cref{sec:schemes,sec:setCover,sec:exactExponentialSoln} are applied by focusing on a single order $a$, letting $q:=|a|$, and renumbering the items in $a$ to be $1,\ldots,q$, ignoring all other items.
Generally in e-commerce fulfillment, $n$ can be much larger than $q$.}
of items $a\subseteq\{1,\ldots,n\}$ that was just purchased.
During each time step, up to one order arrives, which is from region $j$ and is for subset $a$ with probability $\lambda^a_j$, with $\sum_{a,j}\lambda^a_j\le1$.
As is standard in revenue management, we assume a granular division of time such that at most one order can arrive during each time step.
Also, as justified in \citet{jasin2015lp}, we assume that orders cannot contain more than one of any item, and assume a small universe of possible subsets $a$.
We let $\cunit_{kij}$ denote the variable cost of fulfilling one unit of item $i$ from FC $k$ to location $j$,
and let $\cfix_{kj}$ denote the fixed cost of sending a package (containing one or more items) from FC $k$ to location $j$.

The goal is to dynamically decide the FC's to use to fulfill the items in each order that arrives over the time horizon, to minimize total expected cost.
Note that if an FC $k$ is used to fulfill a subset $a'\subseteq a$ of an order from a location $j$, then the cost required to send that package is $\cfix_{kj}+\sum_{i\in a'}\cunit_{kij}$.
All items in each arriving order must be fulfilled from some FC, where we assume the existence of a null FC 0 with infinite inventory so that this is always feasible, with $\cunit_{0ij}$ denoting the ``shortage'' cost of failing to fulfill one unit of item $i$ to region $j$.

\subsubsection*{LP benchmark.}
Solving for the optimal dynamic fulfillment policy using dynamic programming is intractable, since the state space is exponential in the number of items.
Thus, the following ``deterministic'' LP benchmark\footnote{This is identical to the linear program defining $\tilde{J}_{DLP}$ \citep[p.~1340]{jasin2015lp}, except we have let $u^a_{kij}$ and $y^a_{kj}$ represent their variables $U^a_{kij}$ and $Y^a_{kj}$ divided by $T\lambda^a_j$, respectively.} is often used to derive heuristic policies and bound their suboptimality relative to the optimal dynamic programming policy.
\begin{align*}
\DLP:=\min\ \sum_{a,k,j} T\lambda^a_j\left(\sum_{i\in a}\cunit_{kij}u^a_{kij}+\cfix_{kj}y^a_{kj}\right)
\\ \text{s.t. } \sum_j\sum_{a\ni i}T\lambda^a_ju^a_{kij} &\le b_{ki} &\forall k,i
\\ \sum_k u^a_{kij} &=1 &\forall a,j,i\in a
\\ y^a_{kj}\ge u^a_{kij} &\ge0 &\forall a,k,j,i\in a
\end{align*}

In the linear program defining $\DLP$, for any subset $a$ of items ordered from any region $j$, variable $u^a_{kij}$ represents the proportion of times item $i\in a$ should be fulfilled from FC $k$, with constraint $\sum_k u^a_{kij}=1$ for each such item $i$ in the order.
Meanwhile, variable $y^a_{kj}$ represents the probability that a FC $k$ would have to be used at all, which is constrained to be at least $u^a_{kij}$ for any single item $i\in a$.
Note that in an optimal solution we can always assume $y^a_{kj}=\max_{i\in a}u^a_{kij}$ for all $a,k,j$.
These variables $u^a_{kij}$ and $y^a_{kj}$ correspond to our variables $u_{ki}$ and $y_k$ from earlier, where we had dropped scripts $a,j$ to focus on a single multi-item order from a single region.

Moreover, the first constraint enforces that the expected number of times any FC $k$ fulfills any item $i$ (to any region $j$, as part of any subset $a$ containing $i$) does not exceed its starting inventory $b_{ki}$.
Finally, the objective value defining $\DLP$ represents the total expected cost of the LP benchmark over the time horizon, accounting for unit costs, fixed costs, as well as shortage costs (recalling that there is a null FC $k=0$).
This interpretation of $\DLP$ intuitively leads to the following lemma.

\begin{lemma}[\citet{jasin2015lp}] \label{lem:lpUB}
For any instance of the problem, the expected cost paid by any dynamic fulfillment policy must be at least the value of $\DLP$ for that instance.
\end{lemma}

\subsubsection*{Randomized fulfillment algorithm and reduction result.}
In light of the interpretation of the linear program defining $\DLP$ above, \citet{jasin2015lp} also use it to derive the following \textit{randomized fulfillment} heuristic.
First, we solve the LP, hereafter using $u^a_{kij},y^a_{kj}$ to refer to a fixed optimal solution.
At each time step $t=1,\ldots,T$, if an order for subset $a$ comes from region $j$, the heuristic policy randomly chooses an FC $k$ to fulfill each item $i\in a$ according to probabilities $u^a_{kij}$, independently across time steps, without adapting at all to the remaining inventory.
If the chosen FC for an item has stocked out, then that item is simply not fulfilled (i.e.\ the null FC is used).

This randomized fulfillment heuristic that does not rely on real-time inventory information has been shown to perform well asymptotically, although its theoretical guarantee depends on \textit{how} exactly FC's are chosen to fulfill items during each time step, namely, the $\alpha$-competitive rounding scheme that is used.
\citet{jasin2015lp} show that the unit and shortage costs paid by the randomized fulfillment heuristic is asymptotically optimal relative to the
%sum $\sum_{q,k,j}T\lambda^q_j\sum_i\cunit_{kij}u^q_{kij}$ from
$\DLP$, but the bottleneck is the fixed costs, where every time an order for subset $a$ comes from region $j$ (regardless of asymptotics) the cost paid could be $\alpha$ times as much as the $\DLP$.
%the sum $\sum_k T\lambda^q_j\cfix_{kj}y^q_{kj}$ from $\DLP$.
Here $\alpha$ depends on $a$ and $j$, and using the correlated rounding schemes from \Cref{thm:log_n,thm:d} in this paper in conjunction with the results from \citet{jasin2015lp}
we can always guarantee an $\alpha$-competitive rounding scheme where
\begin{align} \label{eqn:bestOf3}
\alpha=\min\Big\{1+\ln(|a|),(\min_{i\in a}\max_k u^a_{kij})^{-1},B(|a|)
\Big\}
\end{align}
and $B(\cdot)$ is the function from \citet{jasin2015lp}.

\citet{jasin2015lp} show that the asymptotic cost paid by the randomized fulfillment heuristic relative to $\DLP$, assuming it chooses the correlated rounding scheme corresponding to the smallest argument in~\eqref{eqn:bestOf3} whenever any subset $a$ is ordered from any region $j$, is a weighted average of expression~\eqref{eqn:bestOf3} over $a$ and $j$.
To formally state this result, we need to finally define what ``asymptotic'' means.
Here, one considers a scaling regime where for any fixed instance and any $\theta\ge0$, the ``scaled instance'' is defined to the the one where the horizon length $T$ has been replaced by $\theta T$ while each starting inventory $b_{ki}$ has also been replaced by $\theta b_{ki}$.  Let $\DLP(\theta)$ denote the optimal objective value $\DLP$ on the instance scaled by $\theta$, and let $\ALG(\theta)$ denote the expected cost paid by the randomized fulfillment heuristic on the same scaled instance.
The following is then implied by the proof of Theorems~1 and~2 from \citet{jasin2015lp} (see \citet[p.~ec5]{jasin2015lp}), combined with our discussion above.
\begin{theorem}[\citet{jasin2015lp}] \label{thm:dynamicFul}
In the multi-item e-commerce fulfillment problem,
\begin{align} \label{eqn:reduction}
\lim_{\theta\to\infty}\frac{\ALG(\theta)}{\DLP(\theta)}\le\frac{\sum_{a,k,j}\lambda^a_j\cfix_{kj}y^a_{kj}\min\Big\{1+\ln(|a|),(\min_{i\in a}\max_{k'} u^a_{k'ij})^{-1},B(|a|)\Big\}}{\sum_{a,k,j}\lambda^a_j\cfix_{kj}y^a_{kj}}.
\end{align}
\end{theorem}

Since any fulfillment policy must pay cost at least $\DLP(\theta)$ by \Cref{lem:lpUB}, this shows that the randomized fulfillment heuristic cannot be worse than the optimal dynamic program by a factor greater than the RHS of~\eqref{eqn:reduction}.
The RHS of~\eqref{eqn:reduction} is a weighted average of the minimum of the guarantees from three different rounding schemes, and was referred to as $\beta$ in the Introduction.
In order to achieve this, the randomized fulfillment heuristic must choose for every incoming order $a$ the rounding scheme with the best guarantee among that of \citet{jasin2015lp}, and our two new ones.
Simpler bounds can also be derived by relaxing the RHS of~\eqref{eqn:reduction}; e.g.
\begin{align} \label{eqn:2089}
\lim_{\theta\to\infty}\frac{\ALG(\theta)}{\DLP(\theta)}
\le
\frac{\sum_{a,k,j}\lambda^a_j\cfix_{kj}y^a_{kj}\Big(1+\ln(|a|)\Big)}{\sum_{a,k,j}\lambda^a_j\cfix_{kj}y^a_{kj}}
\le\frac{\sum_{a,k,j}\lambda^a_j\cfix_{kj}y^a_{kj}\max_{a'}\Big\{1+\ln(|a'|)\Big\}}{\sum_{a,k,j}\lambda^a_j\cfix_{kj}y^a_{kj}}
=1+\ln(\max_{a'}|a'|).
\end{align}

%Our guarantee on the RHS of~\eqref{eqn:reduction} illustrates the power of having different correlated rounding schemes at our disposal for different types of orders that could arrive.
\citet[Thm.~2]{jasin2015lp} prove the same guarantee as \Cref{thm:dynamicFul} except with the $\min\{\cdot\}$ replaced by just $B(|a|)$, while \citet{zhao2020multi} prove the same result where the upper bound on the RHS is 1 (i.e.\ prove asymptotic optimality) if there are only two FC's in the network.
We emphasize that all of these asymptotic guarantees which have eliminated the unit and shortage costs only hold if the LP inventory constraints are satisfied in expectation at every time step, justifying why all of these papers study correlated rounding schemes.

Finally, we argue that the corollary of \Cref{thm:dynamicFul} depicted in~\eqref{eqn:2089}, arising from the $(1+\ln(|a|))$-competitive rounding scheme in our paper, is in fact tight.  In \Cref{sec:setCover} we had already shown that $1+\ln(|a|)$ is the best-possible guarantee for the correlated rounding problem, but here we show computational hardness for dynamic fulfillment, again through a reduction to Set Cover.

\begin{proposition} \label{prop:compHard}
For any positive integer $q$, it is NP-hard to solve dynamic fulfillment using total cost less than $(1-o(1))\ln(q)$ times the optimal cost (given by a computationally-unconstrained dynamic program), even if all orders have size $q$ and even on the scaled instance as $\theta\to\infty$.
\end{proposition}

\proof{Proof of \Cref{prop:compHard}.}
Given any instance of (unweighted) Set Cover, as defined in \Cref{prob:setCover}, we show how it can be represented by an instance of the dynamic fulfillment problem, as defined in this \namecref{sec:multiOrder}.
Recall that $q,K$ were the number of items, sets respectively in the Set Cover problem.
Consider a dynamic fulfillment problem with base time horizon $T=1$, one region, a deterministic order type $a$ of size $q$, and $K$ FC's.
All unit shipping costs are 0 and fixed shipping costs are 1.
Starting inventory $b_{ki}$ equals 1 if $i\in U_k$, and 0 otherwise.
Let $\theta$, the positive integer by which both the time horizon and starting inventories are scaled, be arbitrary.

Due to the inventory configuration, an item $i$ can only be feasibly assigned to an FC $k$ if $i\in U_k$, i.e.\ if item $i$ was covered by set $k$.
The assignment of an item $i$ to any feasible FC $k$ that is already being used (i.e.\ whose fixed shipping cost is being paid) is irrelevant, since all such FC's would start with $\theta$ units of item $i$ and the inventory constraint is not binding.
Therefore, the decision at every period in the dynamic fulfillment problem is identical and equivalent to the minimization problem of the given Set Cover instance, where the goal is to choose a minimum set of FC's to use such that each of the $q$ items in the order can be assigned.  The objective functions also coincide.

Therefore, if it were possible to solve dynamic fulfillment using total cost less than $(1-o(1))\ln(q)$ times the optimum, then it would be possible to solve Set Cover using total cost less than $(1-o(1))\ln(q)$ times the optimum.
By \citet{dinur2014analytical}, the latter statement would imply that P=NP.
Since the scaling parameter $\theta$ was arbitrary, the proof is now complete.
\Halmos\endproof

\begin{remark}
Very recently, \citet{amil2022multi} propose an eye-opening approach to the dynamic fulfillment problem that still uses the randomized fulfillment heuristic but solves a bigger LP that is tighter than $\DLP$.
This bigger LP explicitly models the different ``methods'' by which the items in an order can be split across FC's and fulfilled, obfuscating the need for a correlated rounding scheme.
The authors show that the value of $\ALG(\theta)$ relative to their LP approaches 1 as $\theta\to\infty$, achieving asymptotic optimality
and seemingly contradicting \Cref{prop:compHard}.
However, generally there could be exponentially many ways to split a $q$-item order across $K$ FC's, so without restrictions on the methods, their LP cannot be solved in polynomial time (unless P=NP).
Therefore, in unrestricted settings with large orders, solving the smaller LP and using our correlated rounding procedure is still highly relevant.
\end{remark}

\section{Numerical Study} \label{sec:numExp}

We test our $\alpha$-competitive rounding schemes on the general multi-item dynamic fulfillment problem formalized in \Cref{sec:multiOrder}.
We construct instances aimed to model the operations of a large e-tailer in the continental United States, following the setup of \citet{jasin2015lp} as closely as possible.
%We make our Julia code publicly available so that our numerical study can be replicated.
Our code is in Julia, uses the JuMP \citep{dunning2017jump} package, and is made publicly available at \url{https://github.com/Willmasaur/multi_item_e_commerce_fulfillment}.

\subsubsection*{Regions, fulfillment centers, costs.}
We allow orders to arrive from regions corresponding to the $99$ largest metropolitan areas in the U.S., excluding Honolulu, HI.
The arrival rate from each region is scaled by its 2022 population according to the US Cities Database on \url{https://simplemaps.com/data}.
Meanwhile, we take the $10$ largest Amazon.com Inc. fulfillment centers that were operational\footnote{Compiled from the information at \url{https://www.mwpvl.com/html/amazon_com.html}; available with our code.
} as of 2015 and assume all items are shipped from one of these centralized FC's.
Following \citet{jasin2015lp}, the fixed cost of packaging a box at any FC $k$ for any region $j$ is $\cfix_{kj}=8.759$, while the cost of shipping a single item $i$ from any FC $k$ to any region $j$ is $\cunit_{kij}=0.423+0.000541 \dist_{kj}$, where $\dist_{kj}$ is the air distance between FC $k$ and region $j$ in miles.
Not fulfilling an item $i$ costs double the maximum distance; see \citet{jasin2015lp} for details.

We note that our city populations and FC locations may differ from \citet{jasin2015lp}, as the exact sources they cited are no longer publicly available.
We also procedurally diverge from \citet{jasin2015lp} by always selecting the largest cities and fulfillment centers, whereas they select randomly when fewer than 99 cities or fewer than 10 FC's are needed.  We believe this to generate a more interesting smaller network, because the 10 largest cities are spread out across the corners while the 5 FC's are located in the middle, resulting in difficult fulfillment decisions where a city can be ``nearby'' to multiple FC's (see the data files provided with our code for details).

\subsubsection*{Order types, demand rates, starting inventories.}
Order types $a$ each denote a subset of size up to $\nmax$, from a universe of $n$ items.
For each size in $1,\ldots,\nmax$, there are $\nper$ fixed order types, each of which is a subset of the $n$ items drawn uniformly at random with the correct size.
There is also an order type with size 0, which represent the lack of a customer arrival at a time step.
Note that the total number of order types is $1+\nmax\nper$, which we denote using $Q$.

The demand probabilities are first split randomly between the order sizes $0,1,\ldots,\nmax$, and then for each size, split randomly between the types with that size.
This yields a $Q$-dimensional probability vector, i.e.\ a vector whose entries are non-negative and sum to 1.
Then, a $QJ$-dimensional probability vector is constructed by further splitting each order type among the metropolitan areas according to their populations.
This vector $(\lambda^a_j)_{a,j}$ is then used as input for the dynamic fulfillment problem.

Finally, to determine starting inventories, each FC $k$ first randomly decides whether to carry each item $i$, independently with probability $\pcarry$.
Then, for each region $j$, the closest FC $k$ that carries each item $i$ is identified as $\closest_{i,j}$.
For an item $i$, its ``demand'' at an FC $k$ is
\begin{align*}
\demand_{k,i} = \sum_{a\ni i}\sum_j \bI(\closest_{i,j}=k)\cdot \lambda^a_j,
\end{align*}
where we sum over all queries $a$ containing a copy of item $i$, and consider only the regions $j$ for which FC $k$ is identified as the closest when summing over arrival probabilities $\lambda^a_j$.
Given these values, starting inventories are then placed so that $b_{ki} = T\demand_{k,i} + \zsafety\sqrt{T \demand_{k,i}(1-\demand_{k,i})}$ for all $k$ and $i$, where we note that the total demand for item $i$ closest to FC $k$ over $T$ time steps is Binomially distributed, with mean $T\demand_{k,i}$ and variance $T \demand_{k,i}(1-\demand_{k,i})$.  The formula for $b_{ki}$ is the ideal inventory level to start with according to a Newsvendor model, with safety stock multiplier $\zsafety$ set to 0.5 for all items.

We note that our procedures for randomly generating order types, demand rates, and carrying decisions follow \citet[EC.3]{jasin2015lp} exactly, in which these methods are justified.  The details of these methods can also be found in our code.

\subsubsection*{Algorithms.}
Like \citet{jasin2015lp}, we test the Myopic fulfillment policy as a baseline, which fulfills each item from the closest FC that carries it, not accounting for split orders and minimizing the number of boxes shipped.
We then consider four different algorithms following the randomized fulfillment heuristic described in \Cref{sec:multiOrder}:
\begin{itemize}
\item Indep: Independent Rounding, as described in \citet{jasin2015lp};
\item JS: Correlated Rounding scheme of \citet{jasin2015lp} based on line partitions;
\item Dilate: $(1+\ln(|a|))$-competitive scheme based on dilated opening times (\Cref{sec:log_n});
\item ForceOpen: $d$-competitive scheme based on dilated times and forced openings (\Cref{sec:d}).
\end{itemize}
\citet{jasin2015lp} compare Indep and JS to the Myopic fulfillment policy; we additionally compare our new correlated rounding schemes Dilate and ForceOpen.

\subsection{Experimental Setups} \label{sec:expSetup}
We consider two experimental setups.
First, in \textbf{\Cref{sec:numExpSize}}, we let the number of regions, FC's, items, and time steps be $J=10, K=5, n=20, T=10^5$ respectively.
The queries and starting inventories are generated with $\nmax$ varying in $\{2,5,10\}$, $\nper$ fixed to 5, and $\pcarry$ fixed to 0.75.
We note that when $\nmax=5$ this is exactly the ``base case'' simulated in \citet{jasin2015lp}.
We vary $\nmax$ to see how different rounding schemes handle different order sizes.

We consider a bigger network in \textbf{\Cref{sec:numExpFlex}}, where the number of regions, FC's, and items are $J=99, K=10, n=100$ respectively.
These represent the largest values considered in \citet{jasin2015lp}, and we also increase $T$ to $10^6$ to better capture asymptotic performance.
The queries are generated with $\nmax$ and $\nper$ increased to 10.
Meanwhile, we vary $\pcarry$ in $\{0.25, 0.5, 0.75\}$ to investigate how in a big sparse network with a small value $\pcarry$, the problem can still be easy and in particular ForceOpen can perform well because $d$ is small.

For each experimental setup, we randomly generate 30 instances, and then randomly generate 30 arrival sequences for each instance.
We use the same arrival sequences for every algorithm to minimize the discrepancy caused by variance in arrival sequences.
We fix $\zsafety$ to be 0.5 throughout our experiments.
All of these aspects match what is done in \citet{jasin2015lp}.

\subsection{Performance on Smaller Network with varying Order Size} \label{sec:numExpSize}

We consider the first experimental setup with the smaller network, generating 30 random instances for each value of $\nmax$ in $\{2,5,10\}$.
For each instance, we consider the benchmark $\DLP$ described in \Cref{sec:multiOrder} which is a lower bound on the cost of any fulfillment algorithm.  We draw 30 arrival sequences to test the performance of the 5 specific algorithms discussed earlier, and
compute the average cost of each algorithm over these 30 arrival sequences.  We consider how much greater this average cost is than the value of $\DLP$ for that instance, expressed as a percentage.
The average of these ``loss'' percentages over the 30 instances are then reported in \Cref{tbl:numExpSize}, for each algorithm.
We also report the average runtime\footnote{See our code for the exact timing functions used.  The time (in seconds) was measured on a Dell Latitude 5510 laptop with an Intel(R) Core(TM) i7-10810U CPU @ 1.10GHz processor and 32GB of RAM.} of each algorithm, which we note is the \textit{total} runtime used to evaluate the 30 arrival sequences for an instance, averaged over instances.
Finally, we report for each algorithm the average number of FC's used per order (not counting the ``null'' FC 0).
%We report two additional metrics for each algorithm: the average number of FC's used per order, and the percentage of orders with at least one item that goes unfulfilled.
\begin{table}
\caption{Performance and runtime metrics for the 5 different algorithms under the 3 different values of $\nmax$.  The best (smallest) performances are bolded for each row.}
\label{tbl:numExpSize}
\centering
\begin{tabular}{c|ccccc}
\hline
& Myopic & Indep & JS & Dilate & ForceOpen \\
\hline
$\nmax=2$ Avg. Loss & 4.3\% & 3.1\% & \textbf{2.3}\% & 2.4\% & 2.4\% \\
$\nmax=5$ Avg. Loss & 12.9\% & 14.9\% & 9.3\% & \textbf{8.3}\% & 8.9\% \\
$\nmax=10$ Avg. Loss & 17.7\% & 16.5\% & 11.7\% & \textbf{8.6}\% & 9.4\% \\
\hline
$\nmax=2$ Runtime per Instance & 0.33s & 0.38s & 1.17s & 0.39s & 0.43s \\
$\nmax=5$ Runtime per Instance & 0.52s & 0.59s & 3.37s & 0.62s & 0.72s \\
$\nmax=10$ Runtime per Instance & 0.84s & 1.03s & 9.12s & 1.09s & 1.32s \\
\hline
$\nmax=2$ Avg. FC's per Order & 0.68 & 0.67 & \textbf{0.66} & \textbf{0.66} & \textbf{0.66} \\
$\nmax=5$ Avg. FC's per Order & 1.29 & 1.22 & 1.16 & \textbf{1.15} & 1.16 \\
$\nmax=10$ Avg. FC's per Order & 1.73 & 1.6 & 1.51 & \textbf{1.44} & 1.46 \\
\hline
%$\nmax=2$ Orders w/ Unfulfilled Items & \textbf{1.13}\% & 1.3\% & 1.28\% & 1.29\% & 1.29\% \\
%$\nmax=5$ Orders w/ Unfulfilled Items & \textbf{2.16}\% & 3.9\% & 3.68\% & 3.66\% & 3.65\% \\
%$\nmax=10$ Orders w/ Unfulfilled Items & \textbf{1.93}\% & 3.37\% & 3.24\% & 3.19\% & 3.18\% \\
%\hline
\end{tabular}
\end{table}

\subsubsection*{Observations from results in \Cref{tbl:numExpSize}.}
Our algorithms perform favorably in comparison to Myopic, Indep, and JS.  Indeed, they pay marginally more cost than JS when $\nmax=2$, and overtake JS as soon as $\nmax=5$, i.e.\ orders have sizes between 1 and 5.  This is surprising in that the theoretical guarantee of JS is better for the values of $n$ in this range.  Similar improvements are observed in terms of the average number of FC's used per order.
Also, we note that the losses of 8.3\% and 8.6\% for Dilate are relative to an (unreasonable) LP benchmark which does not face any stochastic fluctuation;
the loss relative to an actual fulfillment policy that can be implemented (e.g., the optimal dynamic programming policy, given the exponential time required to compute it) would be much smaller.
For this reason, we consider the numbers in \Cref{tbl:numExpSize} more useful for comparing algorithms than for evaluating absolute performance.

%Our algorithms also perform better than Indep and JS on the additional metrics---on average, they use fewer FC's per order, and fewer orders go unfulfilled.
%We note that Indep, JS, Dilate, and ForceOpen all cause more unfulfilled items than the Myopic algorithm, since they randomly assign FC's without checking real-time inventory availability.  Although this could be viewed as a weakness of the randomized fulfillment heuristic, it is evident from the overall performance that this weakness is made up for through mitigating split orders.

A further, perhaps more salient feature of our algorithms is their simplicity and interpretability.
As evidenced in our code, the rounding scheme in Dilate (\Cref{alg:1}) takes 10 lines to write, whereas the rounding scheme in JS took us 100 lines.
Also, the average runtime per instance for Dilate is better than JS by a factor of 5--10.
This seemingly innocuous difference on the smaller network becomes more pronounced on the bigger network, as we see next.

\subsection{Performance on Bigger Network with varying Fulfillment Flexibility} \label{sec:numExpFlex}

We consider the second experimental setup described in \Cref{sec:expSetup}.  We report average losses and runtimes for each of the 5 algorithms, in the same way as defined in \Cref{sec:numExpSize}.  We generate 30 random instances for each value of $\pcarry$ in $\{0.25,0.5,0.75\}$ and report the averages in \Cref{tbl:numExpFlex}.

We note that $\pcarry$ is a measure of fulfillment flexibility, in that a higher value of $\pcarry$ leads to more FC's being able to fulfill each item and hence more flexibility in the network.
Generally this results in a harder fulfillment problem, with a larger value of $d$, which we recall denotes the maximum number FC's carrying any item.
A lower value of $\pcarry$, on the other hand, results in a smaller $d$ and a better guarantee for ForceOpen.

\begin{table}
\caption{
%Performance and runtime for the 5 different algorithms under the 3 different values of $$.  The best (smallest) average loss in each row is bolded.
Performance and runtime metrics for the 5 different algorithms under the 3 different values of $\pcarry$.  The best (smallest) performances are bolded for each row.}
\label{tbl:numExpFlex}
\centering
\begin{tabular}{c|ccccc}
\hline
& Myopic & Indep & JS & Dilate & ForceOpen \\
\hline
$\pcarry=0.25$ Avg. Loss & 34.8\% & 10.2\% & 7.6\% & 5.6\% & \textbf{5.3}\% \\
$\pcarry=0.50$ Avg. Loss & 26.7\% & 23.4\% & 17.7\% & \textbf{12.6}\% & 13.1\% \\
$\pcarry=0.75$ Avg. Loss & 22.3\% & 34.2\% & 23.2\% & \textbf{16.1}\% & 17.7\% \\
\hline
$\pcarry=0.25$ Runtime per Instance & 11.23s & 15.43s & 162.31s & 14.24s & 16.88s \\
$\pcarry=0.50$ Runtime per Instance & 11.89s & 18.25s &    162s & 17.01s & 19.25s \\
$\pcarry=0.75$ Runtime per Instance & 13.01s & 19.33s & 169.27s & 18.59s & 22.09s \\
\hline
$\pcarry=0.25$ Avg. FC's per Order & 3.22 & 1.31 & 1.27 & \textbf{1.22} & 1.24 \\
$\pcarry=0.50$ Avg. FC's per Order & 2.4 & 1.98 & 1.87 & \textbf{1.76} & 1.78 \\
$\pcarry=0.75$ Avg. FC's per Order & 1.71 & 1.79 & 1.62 & \textbf{1.50} & 1.53 \\
\hline
%$\pcarry=0.25$ Orders w/ Unfulfilled Items & \textbf{26.1}\% & 49\% & 47.1\% & 46.8\% & 45.5\% \\
%$\pcarry=0.50$ Orders w/ Unfulfilled Items & \textbf{2.63}\% & 11.7\% & 10.7\% & 10.3\% & 10.1\% \\
%$\pcarry=0.75$ Orders w/ Unfulfilled Items & \textbf{2}\% & 4.32\% & 4.02\% & 3.98\% & 3.99\% \\
%\hline
\end{tabular}
\end{table}

\subsubsection*{Observations from results in \Cref{tbl:numExpFlex}.}
In this bigger network which also has larger order sizes,
%up to $\nmax=10$,
all algorithms perform worse.
Myopic performs particularly poorly with large order sizes, because it will likely always split the order (since not all FC's stock all items).
We can see a greater separation between 
the performance of our algorithms, Dilate and ForceOpen, vs.\ the performance of the other algorithms.
And while we had always observed ForceOpen to both be more complex and perform slightly worse than Dilate, we now see that when $\pcarry=0.25$, it in fact performs better.
This is related to its theoretical guarantee---the value of $d$ tends to be smaller when $\pcarry=0.25$, because each item in expectation is carried in only 2.5 FC's.

There is also now a factor-10 speedup in the runtime of our algorithms compared to JS, which means that the time to finish \textit{per instance} is on the order of tens of seconds instead of minutes.

\subsubsection*{Takeaways from numerical study.}
Under the randomized fulfillment heuristic of \citet{jasin2015lp},
one should generally default to Dilate to perform correlated rounding, because it is simple to implement, fast to run, and performs either the best or close to the best across the different setups.
For orders with 2 items, JS may perform slightly better.
In large sparse networks where each item is carried at very few FC's, ForceOpen may perform slightly better.

\section{Conclusion}

We provide the first improvements to the celebrated correlated rounding procedure of \citet{jasin2015lp} for the problem of multi-item e-commerce order fulfillment.
We derive rounding schemes with guarantees of $1+\ln(q)$ and $d$ respectively, where $q$ is the number of items in the order and $d$ is the maximum number of fulfillment centers containing any item.
The first of these guarantees improves the guarantee of $\approx q/4$ from \citet{jasin2015lp} by an order of magnitude, in terms of the dependence on $q$.
We also show both of our guarantees to be tight, by deriving new relationships with the Set Cover problem.
Testing under a realistic setup originated by \citet{jasin2015lp}, we find the improvement provided by our new rounding schemes to in fact be greater than what their theoretical guarantees suggest.

\ACKNOWLEDGMENT{
This research is partially funded by a grant from Amazon.com Inc., which is awarded through collaboration with the Columbia Center of AI Technology (CAIT).
The author thanks anonymous reviewers from several venues---MSOM SIG 2022, ACDA 2023, and EC 2023---for excellent comments that improved the manuscript.
The author also thanks Levi DeValve, Stefanus Jasin, Aravind Srinivasan, Yehua Wei, and Linwei Xin for sharing background information about this problem.
}

\bibliographystyle{informs2014} % outcomment this and next line in Case 1
\bibliography{bibliography} % if more than one, comma separated

\begin{thebibliography}{18}
\providecommand{\natexlab}[1]{#1}
\providecommand{\url}[1]{\texttt{#1}}
\providecommand{\urlprefix}{URL }

\bibitem[{Acimovic \protect\BIBand{} Farias(2019)}]{acimovic2019fulfillment}
Acimovic J, Farias VF (2019) The fulfillment-optimization problem.
  \emph{Operations Research \& Management Science in the age of analytics},
  218--237 (INFORMS).

\bibitem[{Acimovic \protect\BIBand{} Graves(2015)}]{acimovic2015making}
Acimovic J, Graves SC (2015) Making better fulfillment decisions on the fly in
  an online retail environment. \emph{Manufacturing \& Service Operations
  Management} 17(1):34--51.

\bibitem[{Amil et~al.(2022)Amil, Makhdoumi, \protect\BIBand{}
  Wei}]{amil2022multi}
Amil A, Makhdoumi A, Wei Y (2022) Multi-item order fulfillment revisited: Lp
  formulation and prophet inequality. \emph{Available at SSRN 4176274} .

\bibitem[{DeValve et~al.(2021)DeValve, Wei, Wu, \protect\BIBand{}
  Yuan}]{devalve2021understanding}
DeValve L, Wei Y, Wu D, Yuan R (2021) Understanding the value of fulfillment
  flexibility in an online retailing environment. \emph{Manufacturing \&
  Service Operations Management} .

\bibitem[{Dinur \protect\BIBand{} Steurer(2014)}]{dinur2014analytical}
Dinur I, Steurer D (2014) Analytical approach to parallel repetition.
  \emph{Proceedings of the forty-sixth annual ACM symposium on Theory of
  computing}, 624--633.

\bibitem[{Dunning et~al.(2017)Dunning, Huchette, \protect\BIBand{}
  Lubin}]{dunning2017jump}
Dunning I, Huchette J, Lubin M (2017) Jump: A modeling language for
  mathematical optimization. \emph{SIAM review} 59(2):295--320.

\bibitem[{Jasin \protect\BIBand{} Kumar(2012)}]{jasin2012re}
Jasin S, Kumar S (2012) A re-solving heuristic with bounded revenue loss for
  network revenue management with customer choice. \emph{Mathematics of
  Operations Research} 37(2):313--345.

\bibitem[{Jasin \protect\BIBand{} Sinha(2015)}]{jasin2015lp}
Jasin S, Sinha A (2015) An lp-based correlated rounding scheme for multi-item
  ecommerce order fulfillment. \emph{Operations Research} 63(6):1336--1351.

\bibitem[{Lei et~al.(2018)Lei, Jasin, \protect\BIBand{} Sinha}]{lei2018joint}
Lei Y, Jasin S, Sinha A (2018) Joint dynamic pricing and order fulfillment for
  e-commerce retailers. \emph{Manufacturing \& Service Operations Management}
  20(2):269--284.

\bibitem[{Lei et~al.(2021)Lei, Jasin, Uichanco, \protect\BIBand{}
  Vakhutinsky}]{lei2021joint}
Lei Y, Jasin S, Uichanco J, Vakhutinsky A (2021) Joint product framing
  (display, ranking, pricing) and order fulfillment under the multinomial logit
  model for e-commerce retailers. \emph{Manufacturing \& Service Operations
  Management} .

\bibitem[{Motwani \protect\BIBand{} Raghavan(1995)}]{motwani1995randomized}
Motwani R, Raghavan P (1995) \emph{Randomized algorithms} (Cambridge university
  press).

\bibitem[{Raghavan \protect\BIBand{} Tompson(1987)}]{raghavan1987randomized}
Raghavan P, Tompson CD (1987) Randomized rounding: a technique for provably
  good algorithms and algorithmic proofs. \emph{Combinatorica} 7(4):365--374.

\bibitem[{Talluri \protect\BIBand{} Van~Ryzin(2004)}]{talluri2004theory}
Talluri KT, Van~Ryzin G (2004) \emph{The theory and practice of revenue
  management}, volume~1 (Springer).

\bibitem[{Vazirani(2001)}]{vazirani2001approximation}
Vazirani VV (2001) \emph{Approximation algorithms}, volume~1 (Springer).

\bibitem[{Wang et~al.(2022)Wang, Wang, Deng, Cao, \protect\BIBand{}
  Wang}]{Wang22}
Wang Y, Wang X, Deng Y, Cao L, Wang T (2022) Data-driven order fulfillment
  consolidation for online grocery retailing, working Paper.

\bibitem[{Wei et~al.(2021)Wei, Kapuscinski, \protect\BIBand{}
  Jasin}]{wei2021shipping}
Wei L, Kapuscinski R, Jasin S (2021) Shipping consolidation across two
  warehouses with delivery deadline and expedited options for e-commerce and
  omni-channel retailers. \emph{Manufacturing \& Service Operations Management}
  23(6):1634--1650.

\bibitem[{Xu et~al.(2009)Xu, Allgor, \protect\BIBand{} Graves}]{xu2009benefits}
Xu PJ, Allgor R, Graves SC (2009) Benefits of reevaluating real-time order
  fulfillment decisions. \emph{Manufacturing \& Service Operations Management}
  11(2):340--355.

\bibitem[{Zhao et~al.(2020)Zhao, Wang, \protect\BIBand{} Xin}]{zhao2020multi}
Zhao Y, Wang X, Xin L (2020) Multi-item online order fulfillment: A competitive
  analysis. \emph{Chicago Booth Research Paper} (20-41).

\end{thebibliography}

% FOR SUBMISSION
%\ECSwitch
%\ECDisclaimer
%\ECHead{E-Companion}

% FOR SSRN
%\clearpage

% Appendix here
% Options are (1) APPENDIX (with or without general title) or
%             (2) APPENDICES (if it has more than one unrelated sections)
% Outcomment the appropriate case if necessary
%
% \begin{APPENDIX}{<Title of the Appendix>}
% \end{APPENDIX}
%
%   or
%

%\begin{APPENDICES}
%
%
%
%\end{APPENDICES}

%\theendnotes

% References here (outcomment the appropriate case)

% CASE 1: BiBTeX used to constantly update the references
%   (while the paper is being written).
%\bibliographystyle{informs2014} % outcomment this and next line in Case 1
%\bibliography{bibliography} % if more than one, comma separated

% CASE 2: BiBTeX used to generate mypaper.bbl (to be further fine tuned)
%\input{mypaper.bbl} % outcomment this line in Case 2

%If you don't use BiBTex, you can manually itemize references as shown below.

%%%%%%%%%%%%%%%%%
\end{document}